\documentclass{evn2004}
\usepackage{txfonts}
\usepackage{graphicx}
\begin{document}
   \title{The highest redshift radio quasar as seen with the EVN \\ and the VLA}

   \author{S.~Frey\inst{1}
           \and
	   Z.~Paragi\inst{2}
	   \and
	   L.~Mosoni\inst{3}
	   \and
	   L.I.~Gurvits\inst{2}
          }

   \institute{F\"OMI Satellite Geodetic Observatory, P.O. Box 585, H-1592 Budapest, Hungary
         \and
             Joint Institute for VLBI in Europe, Postbus 2, 7990 AA Dwingeloo, The Netherlands
	 \and
             Konkoly Observatory of the Hungarian Academy of Sciences, P.O. Box 67, H-1525 Budapest, Hungary
             }

   \abstract{
We report on European VLBI Network (EVN) and Very Large Array (VLA) observations of SDSS J0836+0054, the most distant radio-loud quasar known at present ($z=5.774$). The source, with a total flux density of 1.1~mJy, shows a compact structure at $\sim10$~mas angular resolution at 1.6~GHz frequency, with no indication of multiple images produced by gravitational lensing within a $4\arcsec \times 4\arcsec$ field above the brightness level of $\sim100$~$\mu$Jy/beam. More recent 5-GHz EVN observations are being analysed now. As a by-product, a two-sided mas-scale jet structure is suspected in one of the phase-reference calibrator quasars, PKS 0837+012.
   }
   \maketitle
%

\section{Introduction}

The object SDSS J0836+0054 is one of the highest redshift quasars known
to date discovered using multicolor imaging data from the Sloan
Digital Sky Survey (SDSS) (Fan et al.\ \cite{Fan01}) with a redshift $z=5.82$.
Recently a more accurate redshift value of $z=5.774$ was determined using near-infrared spectroscopy (Stern et al.\ \cite{Ster03}). The object is one of the twelve $z>5.7$ quasars found in the SDSS to date (Fan et al.\ \cite{Fan01,Fan03,Fan04}). Among these, SDSS J0836+0054 is the only one identified with a radio source.

The estimated mass of the central black hole in SDSS J0836+0054 is
$4.8\times10^{9} M_{\sun}$, assuming that the quasar is emitting at the
Eddington luminosity and its flux is not significantly magnified by beaming or
gravitational lensing (Fan et al.\ \cite{Fan01}).
Simulations predicted that up to one third of $z\sim6$ quasars may be magnified due to gravitational lensing by a factor of ten or more. Therefore their black hole masses are overestimated by the same
factor (Wyithe \& Loeb \cite{Wyi02a,Wyi02b}). The probability estimations strongly depend on the
quasar luminosity function used.

However, there is no evidence so far for strong gravitational lensing (multiple images) in any of the $z>5.7$ SDSS quasars, despite several detection attempts (e.g. Chandra X-ray imaging (Schwartz \cite{Schw02}), Hubble Space Telescope imaging (Richards et al.\ \cite{Rich04}), deep optical imaging of J1044$-$0125 (Shioya et al.\ \cite{Shio02}), EVN radio imaging of J0836+0054 (Frey et al.\ \cite{Frey03}), analysis of the flux distribution of the Ly$\alpha$ emission of J1030+0524 (Haiman \& Cen \cite{Haim02})).
These results suggest that supermassive ($\sim10^9 M_{\sun}$) black holes have already formed at early cosmological epochs, challenging hierarchical structure formation models, and place limit on the slope of the quasar luminosity function at high redshift (e.g. Wyithe \cite{Wyi04}).

\section{EVN observation at 1.6 GHz}

The quasar SDSS J0836+0054 with a flux density of $\sim1$~mJy at 1.4 GHz offered a challenging chance to be detected and imaged with Very Long Baseline Interferometry (VLBI). Any indication of a milli-arcsecond (mas) scale ``core--jet'' radio structure could be considered as a strong support for the existence of a supermassive black hole. According to the current paradigm, the radio emission of an active galactic nucleus (AGN) originates from incoherent synchrotron emission in compact (parsec-scale) jets in the close vicinity of
the central black hole.

SDSS J0836+0054 was observed at 1.6~GHz using ten antennas of the European VLBI Network (EVN) on 2002 June 8.
We selected PKS 0837+012 (J0839+0104; $z=1.123$, Owen et al.\ \cite{Owen95}) as the phase-reference
calibrator. The angular separation between the target and the reference source
is $\sim0\fdg8$. The total observing time on SDSS J0836+0054 was 2.9 hours. A detailed account of the observation and data reduction is given in Frey et al. (\cite{Frey03}).

The quasar is clearly detected with VLBI at 1.6~GHz with a total flux density of 1.1~mJy, at a position of
$\alpha_{\rm{J2000}}=8^{\rm{h}}36^{\rm{m}}43\fs8606$ and
$\delta_{\rm{J2000}}=0{\degr}54{\arcmin}53\farcs232$. Its structure is compact but appears somewhat resolved at a linear resolution of $\sim70$~pc. (Here we assume a flat
cosmological model with $\Omega_{m}=0.3$, $\Omega_{\Lambda}=0.7$ and
$H_{\rm{0}}=65$~km~s$^{-1}$~Mpc$^{-1}$. In this model, 1 mas
angular separation corresponds to a linear separation of 6.25~pc at the distance
of SDSS J0836+0054.) Apparently most if not all of the radio emission is confined to a single compact object within an angular extent of $\sim10$~mas. The quasar is not multiply imaged by gravitational lensing
at the level of brightness ratio $\loa7$ within a $4\arcsec \times 4\arcsec$ field.

\section{EVN observation at 5 GHz}

The compact mas-scale structure of SDSS J0836+0054 seen at 1.6~GHz, and the $\sim0.6$~mJy flux density measured at 5~GHz with the US National Radio Astronomy Observatory (NRAO) Very Large Array (VLA) (Petric et al.\ \cite{Petr03}) prompted us to initiate a follow-up 5-GHz EVN observation. The goals of the project were to obtain spectral information on mas-scale and to refine the astrometric source position.

The 5-GHz phase-referencing experiment took place on 2003 November 4. The Effelsberg (Germany), Hartebeesthoek (South Africa), Medicina (Italy), Nanshan (China), Noto (Italy), Onsala (Sweden), Sheshan (China) and Westerbork (the Netherlands) radio telescopes were used for a total of 7.5 hours, with dual circular polarization where available. The data were recorded with the tape-based Mark IV recording system at most of the stations. Effelsberg recorded with the Mark IV disk-based system. The recording rate was 512 Mbit s$^{-1}$, which resulted in 64~MHz bandwidth per polarisation using two-bit sampling. The correlation took place at the EVN Data Processor at the Joint Institute for VLBI in Europe (JIVE), the Netherlands.

\begin{figure}
\centering
  \includegraphics[clip=,bb=40pt 165pt 600pt 668pt,width=75mm,
angle=0]{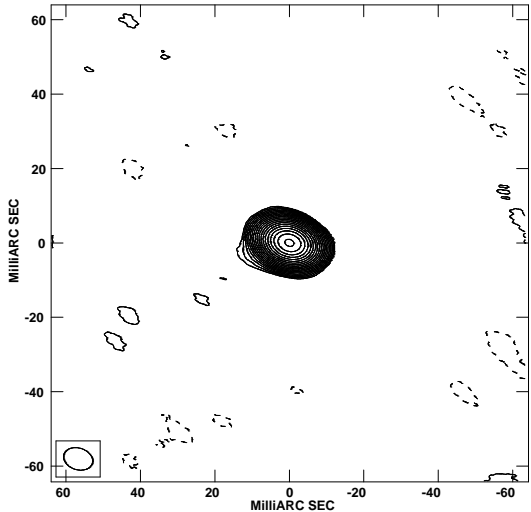}
  \caption{
Naturally weighted 5-GHz VLBI image of the phase-reference calibrator J0836+0052. The positive
contour levels increase by a factor of $\sqrt2$.
The first contours are drawn at $-0.4$ and 0.4~mJy/beam. The peak
brightness
is 219~mJy/beam. The restoring beam is 8.0~mas~$\times$~5.8~mas at a position
angle $72{\degr}$. The Gaussian restoring beam is indicated in the inset. The right ascension ({\it horizontal axis}) and declination ({\it vertical axis}) are relative to the center.
   }
  \label{ref-5GHz}
\end{figure}

In this experiment, we choose J0836+0052 as the phase-reference calibrator source, a quasar with an angular separation of only $\sim5\arcmin$ from SDSS J0836+0054. The reference source proved to be an ideal choice with a slightly resolved, compact structure (Fig.~\ref{ref-5GHz}). The NRAO AIPS package was used for data calibration and imaging. We also included a few scans on the phase-reference calibrator used previously in our 1.6-GHz observation (J0839+0104). This quasar showed relatively complex mas-scale structure at 1.6~GHz, making the relative position determination of SDSS J0836+0054 somewhat uncertain. We will comment on the high resolution radio images of J0839+0104 in Section~\ref{0839}. At the time of writing this report, the analysis of the 5-GHz EVN experiment is not fully completed, because the data on SDSS J0836+0054 are to be re-correlated.

\section{VLA observations}

\begin{figure}
\centering
  \includegraphics[clip=,bb=40pt 165pt 600pt 668pt,width=75mm,
angle=0]{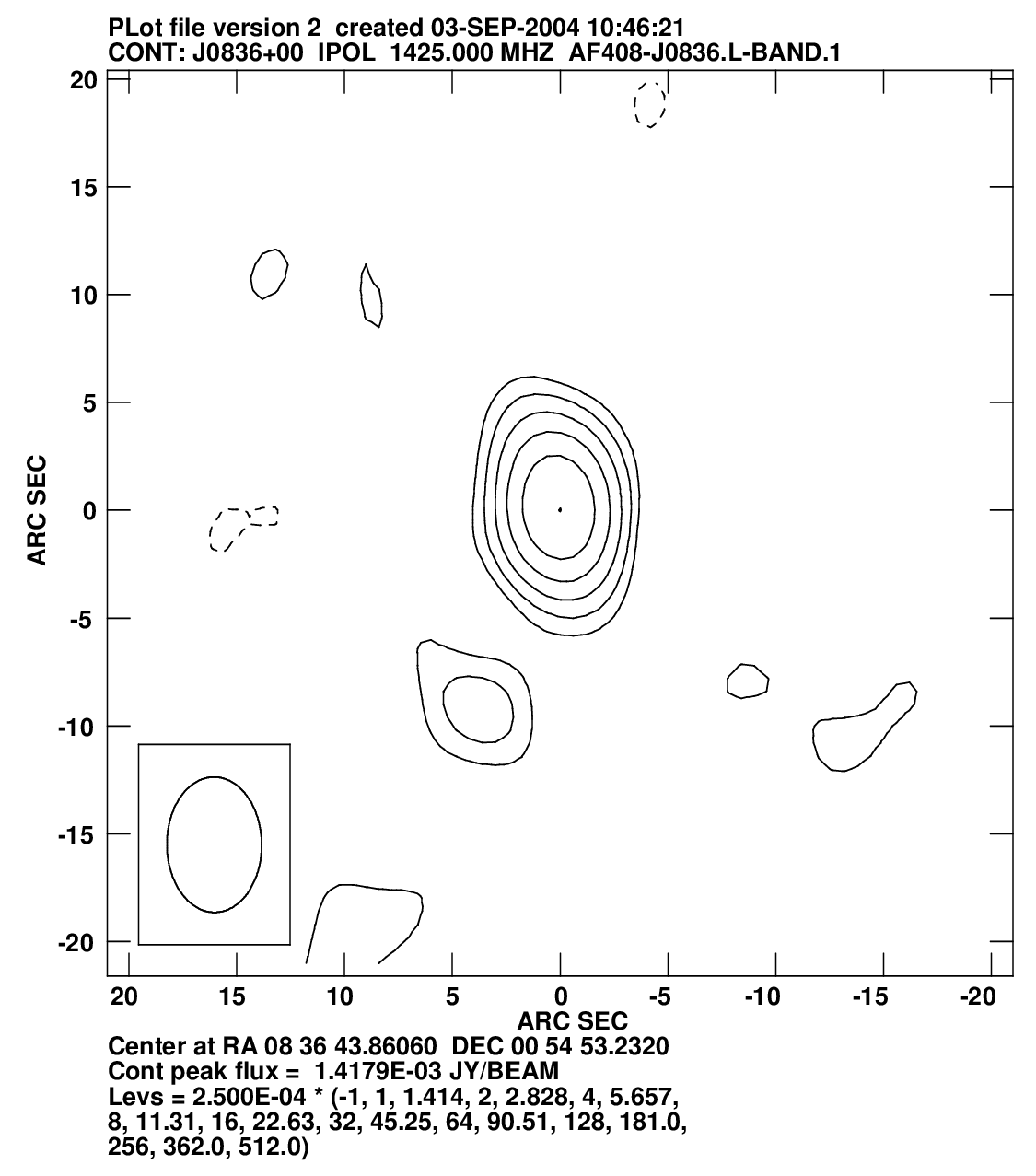}
  \caption{
1.6-GHz VLA image of J0836+0054. The positive
contour levels increase by a factor of $\sqrt2$.
The first contours are drawn at $-0.25$ and 0.25~mJy/beam. The peak
brightness
is 1.418~mJy/beam. The restoring beam is $6\farcs3 \times 4\farcs4$ at a position
angle $1{\degr}$.
   }
  \label{VLA-L}
\end{figure}

\begin{figure}
\centering
  \includegraphics[clip=,bb=40pt 165pt 600pt 668pt,width=75mm,
angle=0]{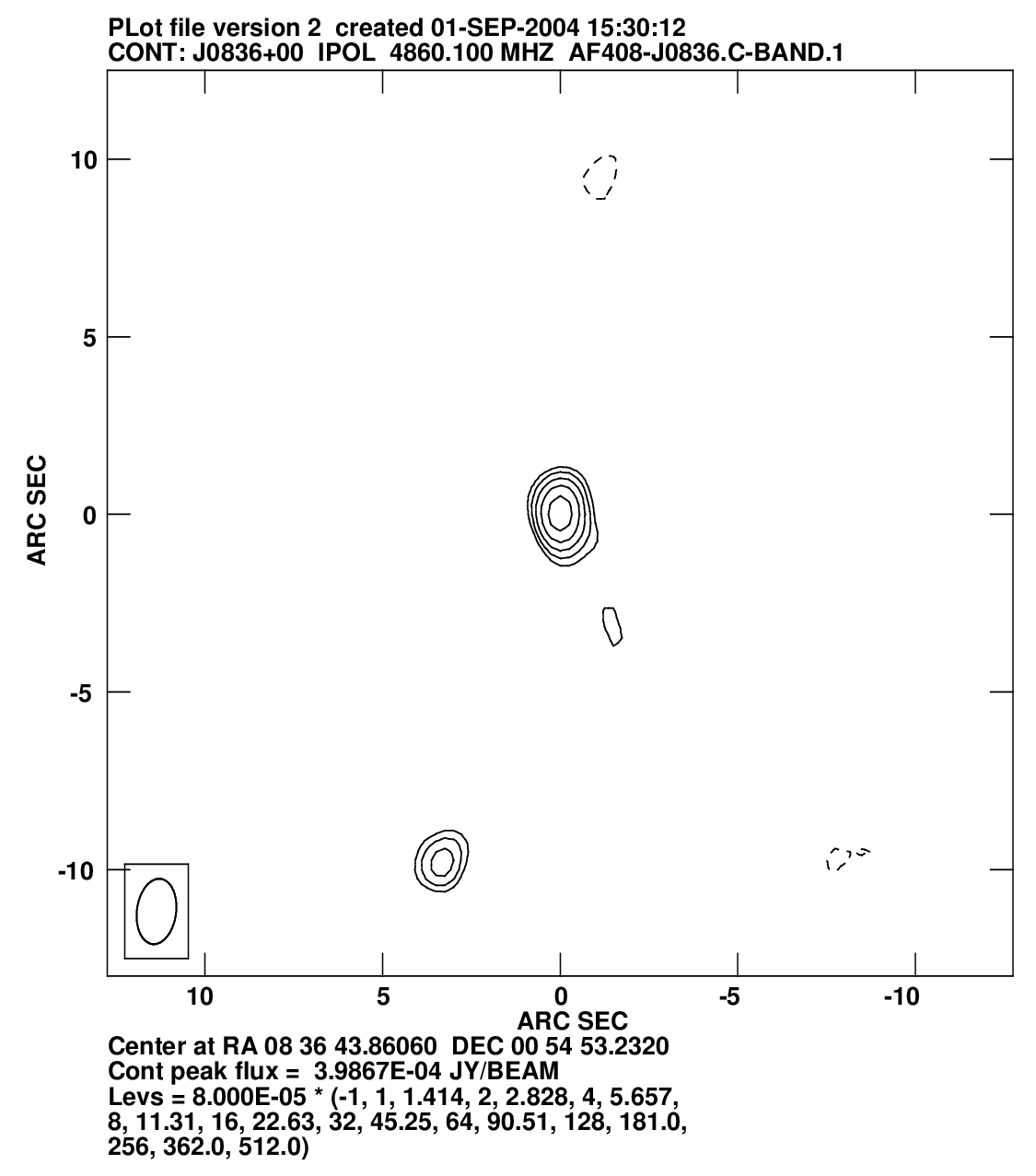}
  \caption{
5-GHz VLA image of J0836+0054. The positive
contour levels increase by a factor of $\sqrt2$.
The first contours are drawn at $-0.08$ and 0.08~mJy/beam. The peak
brightness
is 0.399~mJy/beam. The restoring beam is $1\farcs9 \times 1\farcs1$ at a position
angle $-7{\degr}$.
   }
  \label{VLA-C}
\end{figure}

Sensitive multi-frequency VLA observations of SDSS J0836+0054 were made earlier by Petric et al. (\cite{Petr03}), indicating a steep radio spectrum between 1.4 and 5~GHz (spectral index $\alpha=-0.8$; $S\propto\nu^{\alpha}$). In order to assess the fraction of the flux density originating from the mas-scale compact core at 5~GHz, we conducted short supporting VLA observations of SDSS J0836+0054. The VLA data were taken on 2003 November 5, the day following the 5-GHz EVN experiment. The array was in its B configuration. The flux densities measured ($1.96\pm0.31$~mJy and $0.43\pm0.06$~mJy at 1.4 and 5~GHz, respectively) are consistent with the values obtained by Petric et al. (\cite{Petr03}). Our VLA images (Fig.~\ref{VLA-L} and \ref{VLA-C}) were made using the standard procedures in the NRAO AIPS package. The source SDSS J0836+0054 was observed for 18.5 and 19.5 minutes at 1.4 and 5 GHz, respectively.

A weak nearby radio source located at $9\farcs7$ to the S and $3\farcs4$ to the E of SDSS J0836+0054 is seen in both images. This ``companion'' has been investigated as a possible gravitationally lensed image of the extremely distant quasar. The lensing hypothesis has been ruled out on the basis of different radio spectral indices and a deep optical image (D. Rusin \& B. McLeod, priv. comm.). The latter shows the second source significantly resolved, suggesting that it is associated with a lower redshift galaxy.

Although the possible image separations are expected around $1\arcsec$ or smaller (Richards et al.\ \cite{Rich04}), there are now a few examples known for extremely large separation (up to almost $15\arcsec$) gravitatonal lenses (e.g. Inada et al.\ \cite{Inad03}). The radio spectral slopes of the two sources are similar within the errors based on our limited VLA data. We are not aware of any spectroscopic observation that would ultimately prove that the secondary object is indeed in the foreground.

\section{Comments on the phase-reference calibrator J0839+0104}
\label{0839}

\begin{figure}
\centering
  \includegraphics[clip=,bb=40pt 165pt 600pt 668pt,width=75mm,
angle=0]{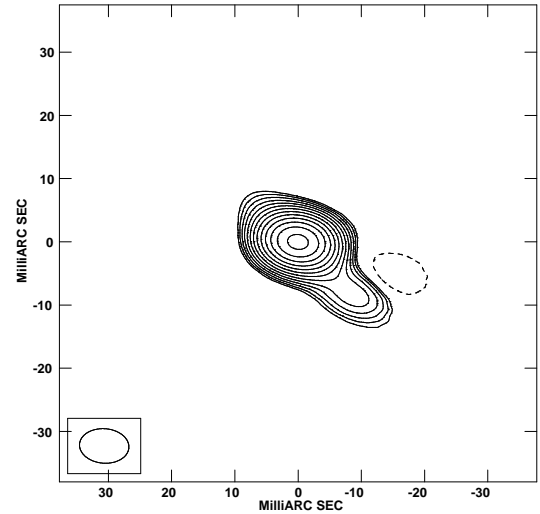}
  \caption{
Naturally weighted 5-GHz EVN image of J0839+0104. The positive
contour levels increase by a factor of $\sqrt2$.
The first contours are drawn at $-0.4$ and 0.4~mJy/beam. The peak
brightness
is 288~mJy/beam. The restoring beam is 7.8~mas~$\times$~5.5~mas at a position
angle $86{\degr}$.
   }
  \label{0839-EVN-nat}
\end{figure}

\begin{figure}
\centering
  \includegraphics[clip=,bb=40pt 165pt 600pt 668pt,width=75mm,
angle=0]{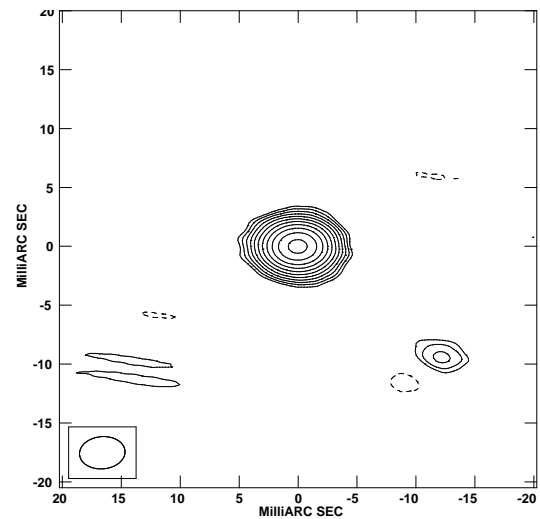}
  \caption{
5-GHz EVN image of J0839+0104 made with {\sc robust=0} weighting parameter in the AIPS task IMAGR. The positive
contour levels increase by a factor of $\sqrt2$.
The first contours are drawn at $-0.7$ and 0.7~mJy/beam. The peak
brightness
is 252~mJy/beam. The restoring beam is 3.9~mas~$\times$~2.7~mas at a position
angle $-85{\degr}$.
   }
  \label{0839-EVN-rob0}
\end{figure}

The high resolution radio structure of the phase-reference calibrator quasar J0839+0104 in our 1.6-GHz EVN experiment shows extended features on both sides of the compact core (Fig.~1 in Frey et al.\ \cite{Frey03}). There is a low surface brightness jet to the SW seen in the naturally weighted image, extending to $\sim30$~mas. If uniform weighting is employed, the improved angular resolution reveals another component at 6.6~mas to the E of the core. In our 5-GHz EVN experiment, the source was observed for a total of 16 minutes to check whether the brightest component in the 1.6-GHz images is the flat-spectrum radio core. Here we present two 5-GHz images (Fig.~\ref{0839-EVN-nat} and \ref{0839-EVN-rob0}) made with different weighting schemes in the AIPS task IMAGR. The extended structure in the SW direction is clearly seen in the naturally weighted image (Fig.~\ref{0839-EVN-nat}). There is also a hint on an extension to the E (see Fig.~\ref{0839-EVN-rob0}), in a similar position angle that was found at 1.6~GHz. This latter image was made with the {\sc robust=0} weighting parameter in IMAGR, in order to closely match in angular resolution the uniformly weighted 1.6-GHz image (Frey et al.\ \cite{Frey03}).

Based on the images made at the two different frequencies, we suspect that the quasar J0836+0104 exhibits a rather unusual two-sided jet structure at mas-scale. The overwhelming majority of radio quasars have one-sided inner radio jets, although there are rare examples of two-sided structures (e.g. Taylor \& Vermeulen \cite{Tayl97}; Mantovani et al.\ \cite{Mant02}). The generally observed asymmetry is due to bulk relativistic motion which boosts the radiation of one of the two intrinsically symmetrical jets. The source J0836+0104 could certainly be a target of more sensitive multi-frequency VLBI imaging to study the nature of its odd structure.

\begin{figure}
  \includegraphics[clip=,bb=473pt 45pt 98pt 748pt,width=45mm,
angle=90]{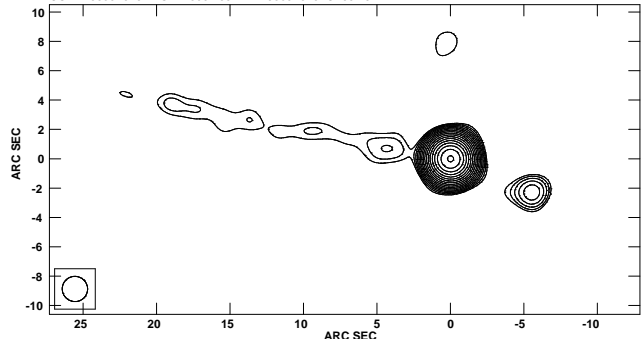}
  \caption{
5-GHz VLA image of J0836+0104, using data on the shortest ($\leq 10^5 \lambda$) baselines only. The positive
contour levels increase by a factor of $\sqrt2$.
The first contours are drawn at $-0.09$ and 0.09~mJy/beam. The peak
brightness
is 240~mJy/beam. The restoring beam is $1\farcs7 \times 1\farcs7$.
   }
  \label{VLA-0839}
\end{figure}

Remarkably, a 5-GHz image (Fig.~\ref{VLA-0839}) made using data obtained from the VLA archive (experiment AR493, observed on 2002 May 5, configuration A) suggests that the two-sided jet structure remains visible out to the $10\arcsec$-scale.

\section{Conclusions}

SDSS J0836+0054, the most distant radio-loud quasar known to date ($z=5.774$) is clearly detected with EVN observations at 1.6 GHz in 2002. The radio structure is compact at $\sim10$ mas angular scale, with a flux density of 1.1~mJy, close to the total flux density value. We found that the source is unlikely to be multiply imaged by gravitational lensing. The 5-GHz EVN observation of the quasar made in 2003 is being processed now. Nearly contemporaneous VLA observations confirm the steep radio spectrum of the source. The flux density measured with the VLA (0.43~mJy), and the compact structure make the VLBI detection probable at 5~GHz.

Analysis of the EVN images of the phase-reference calibrator, the flat-spectrum quasar PKS 0837+012 (J0839+0104, $z=1.123$) made at both 1.6 and 5 GHz frequencies suggests a two-sided mas-scale jet structure, supported by an arcsecond-resolution VLA image at 5 GHz. This is almost unique among radio-loud quasars. It makes this source attractive for more detailed studies in the future.

\begin{acknowledgements}
We thank David Rusin and Brian McLeod for informing us about their optical observations of SDSS J0836+0054.
This research was supported by the Hungarian Scientific Research Fund (OTKA, grant T046097) and the European Commission's I3 Programme RadioNet, under contract No.\ 505818. SF acknowledges the Bolyai Research Scholarship received from the Hungarian Academy of Sciences.
The European VLBI Network is a joint facility of European, Chinese,
South African and other radio astronomy institutes funded by their
national research councils.
The National Radio Astronomy Observatory is a facility of the National Science Foundation opearted under cooperative agreement by Associated Universities, Inc.
This research has made use of the NASA/IPAC Extragalactic Database (NED) which
is operated by the Jet Propulsion Laboratory, California Institute of
Technology, under contract with the National Aeronautics and Space
Administration.
\end{acknowledgements}

\end{document}